\documentclass[twocolumn,showpacs,superscriptaddress,preprintnumbers,amsmath,amssymb,prl]{revtex4}
\usepackage{graphicx}
\usepackage{dcolumn}
\usepackage{color}
\usepackage{bm}
\begin{document}

\title {Large Magnetic Moments of Arsenic-Doped Mn Clusters and their Relevance to 
Mn-Doped III-V Semiconductor Ferromagnetism}
\author{Mukul Kabir}
\altaffiliation{{\it Corresponding author}}
\affiliation{S.N. Bose National Centre for Basic Sciences, JD Block, Sector III, Salt Lake, Kolkata - 700 098, India }
\author{D. G. Kanhere}
\affiliation{Department of Physics and Centre for Modelling and Simulation, University of Pune, Pune - 411 007, India }
\author{Abhijit Mookerjee}
\affiliation{S.N. Bose National Centre for Basic Sciences, JD Block, Sector III, Salt Lake, Kolkata - 700 098, India }
\date{\today}

\begin{abstract}
We report electronic and magnetic structure of arsenic-doped manganese clusters from density-functional 
theory using generalized gradient approximation for the exchange-correlation energy. We find that arsenic 
stabilizes manganese clusters, though the ferromagnetic coupling between Mn atoms are found only in 
Mn$_2$As and Mn$_4$As clusters with magnetic moments 9 $\mu_B$ and 17 $\mu_B$, respectively. For all 
other sizes, $x=$ 3, 5-10, Mn$_x$As clusters show ferrimagnetic coupling. It is suggested that, if grown 
during the low temperature MBE, the giant magnetic moments due to ferromagnetic coupling in Mn$_2$As and 
Mn$_4$As clusters could play a role on the ferromagnetism and on the variation observed in the Curie 
temperature of Mn-doped III-V semiconductors.  
\end{abstract}
\maketitle

Manganese is a unique 3$d$ transition metal element due to its unusual and fascinating 
electronic and magnetic behavior as atom, cluster, crystal and as well as impurity. 
The filled 4$s$ and half-filled 3$d$ shell and the large energy gap $\sim$ 8 eV between 
them prevent significant $s-d$ hybridization and therefore they do not bind strongly 
as Mn atoms begin to form cluster\cite{kabir, mnpre}.
Similarly, $\alpha$-Mn, the most stable form of bulk Mn, has the
least binding energy among all the 3$d$ transition metal elements.
The magnetic properties of Mn clusters are particularly interesting. An early electron 
spin resonance study on small Mn clusters in an inert matrix suggested ferromagnetic ordering 
with magnetic moment 5 $\mu_B$/atom \cite{zee}. More recently, through Stern-Gerlach (SG) 
molecular beam experiment \cite{mark}, Knickelbein found that manganese clusters in the size range 
Mn$_{5}$ - Mn$_{99}$ display ferrimagnetic ordering with a maximum magnetic moment $\sim$ 1.72 
$\mu_B$/atom for Mn$_{12}$, despite of the fact that no known bulk phase of Mn 
displays such ordering. Recent density functional theory (DFT) calculations 
\cite{kabir, mnpre} confirm this ferrimagnetic coupling in the size range 
$x \ge$5 in Mn$_x$ clusters.  
                                                                                                          
Manganese doped semiconductors, such as (GaMn)As and (InMn)As have attracted
considerable attention because of their carrier induced ferromagnetism \cite{ohno}. The Mn dopants 
in these III-V dilute magnetic semiconductors serve the dual roles of provision of
magnetic moments and  acceptor production. Earlier experimental results on 
Ga$_{1-x}$Mn$_x$As indicate a nonmonotonic behavior of Curie temperature $T_c(x)$, 
first increases with the Mn concentration $x$, reaching a maximum of 110 K for 
$x\sim$5\% and then decreases with the further increase of $x$. However, recent 
experimental studies \cite{hightc, hightc1}, under carefully controlled growth and 
annealing conditions, suggest that the {\it metastable} nature and high {\it defect} 
content of low temperature molecular beam epitaxy (MBE) grown Ga$_{1-x}$Mn$_x$As may be 
playing an important role in determining the magnetic properties. In particular, under 
suitable conditions, careful annealing could lead to an enhancement in $T_c$ with 
increasing Mn content \cite{hightc, hightc1}. Chiba {\it et al.} \cite{hightc} have 
reported a maximum $T_c$ as high as 160 K, for 7.4 \% Mn concentration, in a layered 
structure. However, the detailed microscopic nature and the role of the 
{\it metastable defects} in Ga$_{1-x}$Mn$_x$As are not yet well understood both 
theoretically and experimentally and demands more investigation. Strong segregation tendency of the
doped transition metal atoms into the semiconductor host \cite{segregation} makes Mn clustering, 
around As in Ga$_{1-x}$Mn$_x$As samples, a very important issue.

In this Letter, we address the possibility of such clustering of Mn around As, and if so, then 
what is the nature of Mn-Mn magnetic coupling? We have, indeed, found that the binding energy of Mn 
clusters are
substantially enhanced by single As doping by having their hybridized $s-d$ electrons bond
with $p$ electrons of As. This {\it stabilization} is accompanied by the {\it ferromagnetic}
or {\it ferrimagnetic} coupling between Mn atoms, all
of which results in the large cluster magnetic moments, which could play a crucial role in the
observed ferromagnetism and determining the Curie temperature $T_c$, in Mn-doped GaAs and InAs.

Calculations have been carried out using DFT, within the plane wave 
method \cite{pseudo}. We have employed projector augmented-wave method \cite{paw} and 
Perdew-Burke-Ernzerhof exchange-correlation functional \cite{pbe} for the spin-polarized 
generalized gradient approximation (GGA), as implemented in the VASP package \cite{vasp}. The wave functions 
are expanded in a plane wave basis set with the kinetic energy cutoff equal to 337.3 eV 
 and the calculations were carried out at the $\Gamma$ point.  The 3$d$, 
4$s$ for Mn and 4$s$, 4$p$ orbitals for As were treated as valence states. Symmetry unrestricted 
geometry optimizations were performed using quasi Newtonian and conjugate gradient methods until all the 
force components are less than 0.005 eV/\r{A}. Simple cubic supercells are used with neighboring 
clusters separated by at least 12\r{A}  vacuum regions. Several initial structures were studied to 
ensure that the globally optimized geometry does not correspond to the local minima, as well as, for 
all clusters, we have explicitly considered {\it all possible} spin multiplicities to determine the 
ground state magnetic moment.

\begin{figure}[b]
{\rotatebox{0}{\resizebox{4.2cm}{3.7cm}{\includegraphics{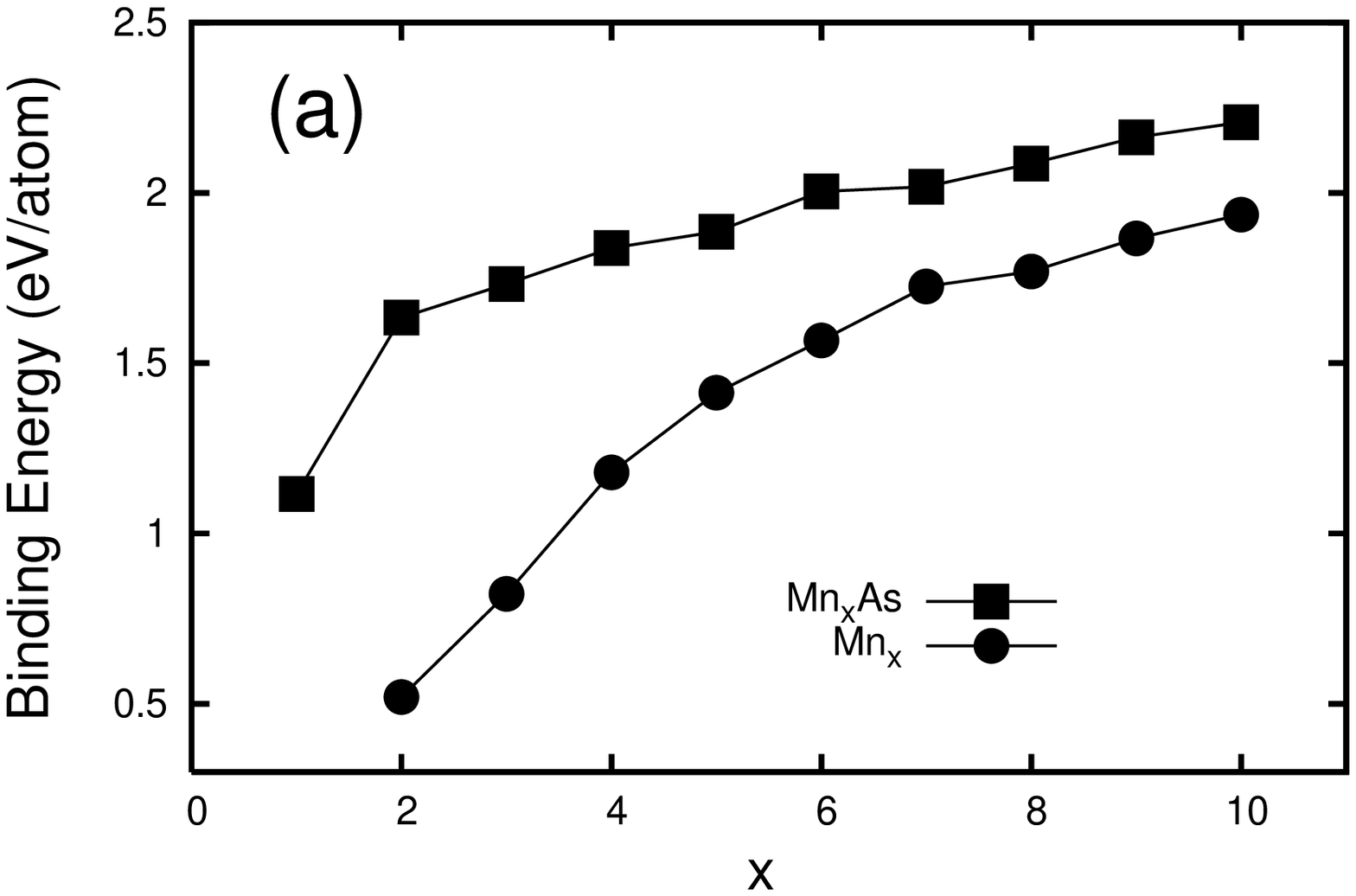}}}}
{\rotatebox{0}{\resizebox{4.2cm}{3.7cm}{\includegraphics{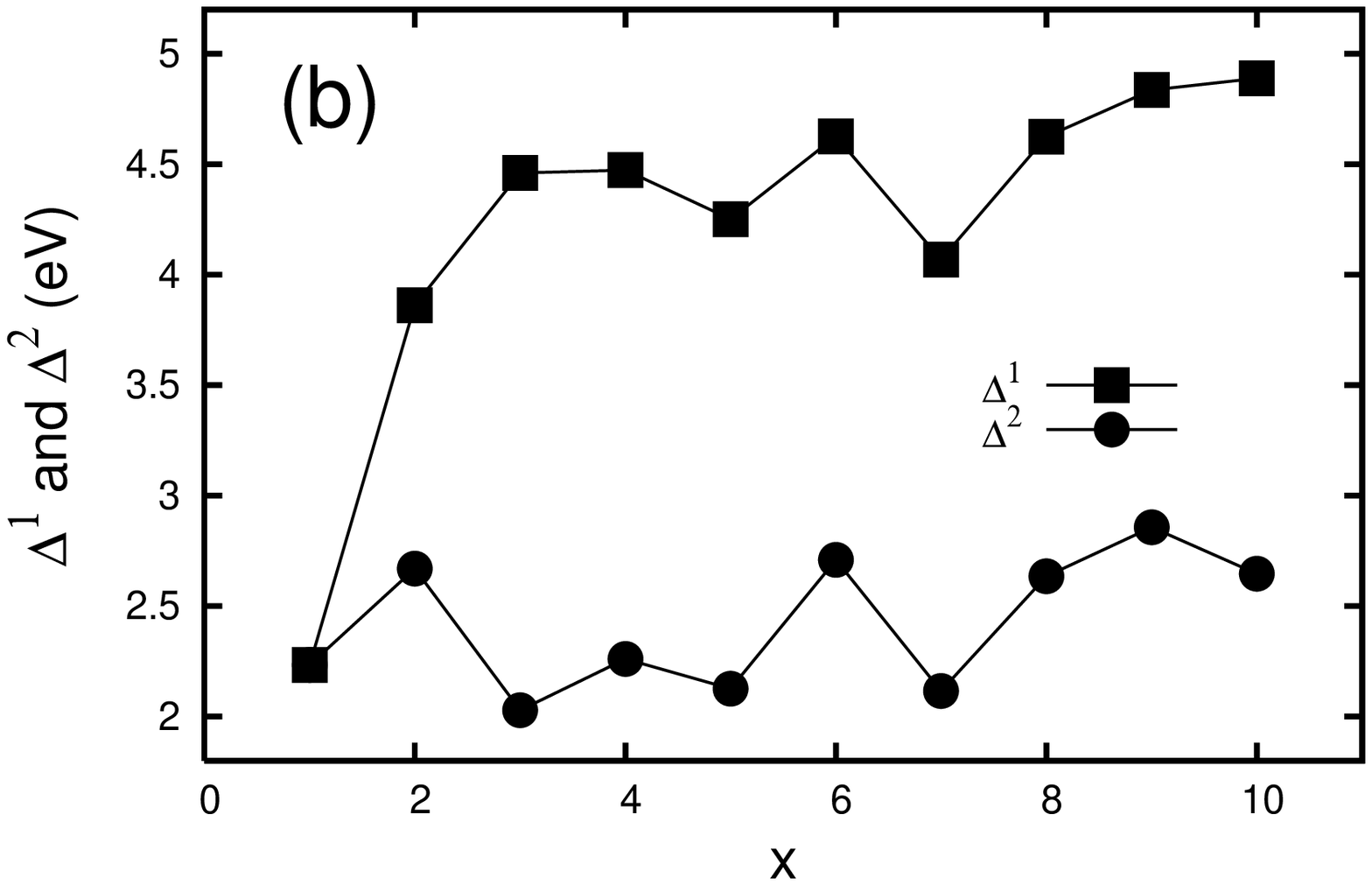}}}}
\caption{\label{fig:energygain} (a) Plot of binding energy/atom with $x$, for Mn$_x$ and Mn$_x$As. 
(b) Two energy gains, $\Delta^1$ and $\Delta^2$, are plotted with $x$. These energy gains are defined as, 
$\Delta^1$ = $-$ [$E$(Mn$_x$As) $-$ $E$(Mn$_x$) $-$ $E$(As)] and $\Delta^2$ = $-$ [$E$(Mn$_x$As) $-$ 
$E$(Mn$_{1-x}$As) $-$ $E$(Mn)].}
\end{figure}

We begin our discussions with pure Mn$_x$ ($x\le$10) clusters, some of which ($x\le$7) have also been
studied previously within the all-electron calculation and different levels of GGA \cite{mnpre}. 
We have found 
ferromagnetic coupling for Mn$_2$, Mn$_3$ and Mn$_4$ with magnetic moment 5 $\mu_B$/atom, the Hund's 
rule value for the free atom. Emergence of ferrimagnetic coupling starts form Mn$_5$. The total magnetic 
moments of Mn$_x$ clusters corresponding to the ground state geometries, for $x$ = 5-10,
 are 3, 8, 5, 8, 7 and 14 $\mu_B$ 
respectively (see Table \ref{tab:magmom}), which are in excellent agreement with the recent 
SG experiment \cite{mark}.
Because of the half-filled 3$d$ and filled 4$s$ shell and a substantial 
gap between them, we found Mn$_2$ is a weakly bound van der Walls (vdW) dimer with  bond length 2.58 \r{A} and 
binding energy 0.53 eV/atom, as it is evident from the low experimental value \cite{morse}.
Antiferromagnetic Mn$_2$ dimer is 0.53 eV 
higher in 
energy, whereas for Mn$_3$, the frustrated antiferromagnetic structure is almost degenerate (0.05 eV 
higher in energy) with the ferromagnetic structure. 
As $x$ increases, binding energy increases monotonically [Fig.\ref{fig:energygain}(a)], reaches a 
value 1.94 eV/atom, 66\% of bulk $\alpha$-Mn, for Mn$_{10}$. We note, generally, the Mn-Mn bonds
between like spins are larger (2.30-2.85 \r{A}) than that of between opposite spins (2.20-2.75 \r{A}). 
Several isomers lying close in energy to 
the ground state of Mn$_x$ were found which will be discussed elsewhere \cite{kabir}, in detail.

In the Fig.\ref{fig:structure}, we depict the ground state geometries of Mn$_x$As clusters with their
spin ordering. The MnAs dimer has much higher binding energy, 
1.12 eV/atom, and much shorter bond length, 2.21 \r{A}, than the Mn$_2$ vdW dimer. 
We have repeated the calculations of MnAs dimer, including  Mn 3$p$ as valence state \cite{repeat}, and have 
obtained an optimized bond length 2.22 \r{A} and binding energy 1.08 eV/atom with same total magnetic 
moment. 
Binding energy increases substantially to 1.63 eV/atom for the isosceles triangular Mn$_2$As. 
The Mn-Mn distance, 2.59 \r{A}, in Mn$_2$As is almost 
equal to the Mn-Mn distance in Mn$_2$, 2.58 \r{A}, and Mn-As-Mn bond angle is found to be 68$^0$. 
As more Mn atoms are added, the structures take three-dimensional shape, and the determination of 
ground state become a delicate task, as more than one geometric and/or magnetic isomers, very close
in energy to the ground state are found, which  will be presented elsewhere in detail. As mentioned earlier, 
this was done by minimizing the total energy of each cluster with respect to {\it all possible} spin 
multiplicities for every geometric structure studied. The presence of an As atom makes Mn$_3$As 
cluster tetrahedral, however, Mn$_4$As is a Mn$_4$ tetrahedron with As at a face cap. Mn$_6$As is 
the only cluster whose geometry differs significantly from a pure Mn$_6$ cluster. Mn$_6$ is 
octahedral, whereas Mn$_6$As is a pentagonal bipyramid, where the As atom is trapped in the pentagonal
ring. As $x$ in Mn$_x$As increases, binding energy increases very slowly from 1.63 eV/atom  
for Mn$_2$As and tends to saturate to a value 2.21 eV/atom for Mn$_{10}$As (Fig.\ref{fig:energygain}(a)). 
The {\it shortest} Mn-As bond 
length increases from 2.21 \r{A} for MnAs to 2.46 \r{A} for Mn$_{10}$As, which is 4\%  and 2\% shorter than 
the Mn-As distance in $\alpha$-MnAs and Ga$_{1-x}$Mn$_x$As, respectively, whereas the  {\it shortest} Mn-Mn 
distance decreases from 2.59 \r{A} for Mn$_2$As to 2.23 for Mn$_{10}$As. Generally, we find the same trend
as seen in the pure Mn$_x$ clusters that the bonds 
between Mn atoms of opposite spin to be somewhat shorter (2.20-2.60 \r{A}) than the bonds between Mn atoms
of like spin (2.50-2.90 \r{A}), whereas all Mn-As distances vary between 2.20-2.60 \r{A}. All the Mn-As-Mn
bond angles in these clusters vary in between $\sim$ 60-70$^0$. All the clusters in the 
Fig.\ref{fig:structure} and their respective isomers are magnetically stable i.e. both the spin gaps are
positive: the lowest unoccupied molecular orbital of the minority (majority) spin lies above the highest
occupied molecular orbital of the majority  (minority) spin. These two spin gaps, $\delta_1$ and 
$\delta_2$ \cite{delta}, for Mn$_2$As (0.83 and 1.34 eV) and Mn$_4$As (0.89 and 1.14 eV) are the highest
among all clusters. As $x$ increases, $\delta_1$ and $\delta_2$ decrease to a value 0.47 and 0.35 eV, 
respectively, for Mn$_{10}$As.

The next important issue is to see whether these Mn clustering around single As are at all 
energetically favorable or not. To understand this point, we calculate two different energy gains, 
$\Delta^1$ - the energy gain in adding an As atom to a Mn$_x$ cluster and $\Delta^2$ - the energy 
gain in adding a Mn atom to a Mn$_{1-x}$As cluster. Fig.\ref{fig:energygain}(a) shows that due to the
lack of hybridization between the 4$s$ and the 3$d$ electrons binding energy of pure Mn$_x$ clusters 
are very small. However, as an As atom is attached, the 4$s^2$ electrons of Mn interact with the 
4$p^3$ electrons of As, which results in the substantial enhancement 
in the bonding (Fig.\ref{fig:energygain}(a)), and consequently, $\Delta^1$ increases with $x$, which 
finally tends to saturate 
(Fig.\ref{fig:energygain}(b)).  $\Delta^2$ gives the number that how many Mn atoms can be bonded to a 
single As atom, which is still significant, 2.65 eV, for Mn$_{10}$As. These behaviors of $\Delta^1$ and
$\Delta^2$ indicate that the Mn clusters around As are energetically favorable and we, therefore, argue that 
they are, likely to be, present in the low temperature MBE grown (GaMn)As/(InMn)As.

\begin{table}
\caption{\label{tab:magmom}Total cluster magnetic moments $\mu_x$ of pure Mn$_x$ and Mn$_x$As clusters,
corresponding to the ground state, for cluster size $x \le 10$.}
\begin{tabular}{cccccc}
\hline
\hline
$x$ \phantom{xx}& \multicolumn{2}{c}{ $\mu_x $($\mu_B$)} & \phantom{xxxx} $x$ \phantom{xx} & \multicolumn{2}{c} {$\mu_x$($\mu_B$)} \\
    \phantom{xx}&    Mn$_x$  & Mn$_x$As  & \phantom{xxxx}    \phantom{xx} &    Mn$_x$  & Mn$_x$As   \\ 
\hline
1  \phantom{xx}&    5   &  4   & \phantom{xxxx}  6 \phantom{xx} &  8  & 9   \\
2 \phantom{xx} &    10  &  9   & \phantom{xxxx}  7  \phantom{xx}&  5  & 6   \\
3  \phantom{xx}&    15  &  4   & \phantom{xxxx}  8  \phantom{xx}&  8  & 7   \\
4  \phantom{xx}&    20  &  17  & \phantom{xxxx}  9  \phantom{xx}&  7  & 10  \\
5  \phantom{xx}&    3   &  2   & \phantom{xxxx}  10 \phantom{xx}&  14 & 13  \\
\hline
\hline         
\end{tabular}
\end{table}

Now we ask the next and most important question that what is the Mn-Mn magnetic coupling in these
Mn$_x$As clusters? The total magnetic moments of Mn$_x$As clusters corresponding to the ground state 
geometries are given in Table \ref{tab:magmom}. These large magnetic 
moments generally arise from the ferrimagnetic coupling between the moments at Mn sites with the exceptions
for Mn$_2$As and Mn$_4$As, where the magnetic coupling is ferromagnetic. We note, generally, no change 
in the nature of magnetic coupling between the Mn sites in Mn$_x$As clusters from their respective pure 
Mn$_x$. However, only
for Mn$_3$As, the Mn-Mn coupling behavior changes form ferromagnetic to ferrimagnetic, due to As doping, where 
a ferromagnetic tetrahedral structure with total moment 12 $\mu_B$ is found to be 0.12 eV higher. 
The nature of the magnetic coupling is clear from their respective constant spin density surfaces, which 
are plotted for Mn$_2$As, Mn$_3$As, Mn$_4$As and Mn$_8$As in the Fig.\ref{fig:densities}.
The ground state magnetic moments of Mn$_x$As, for $x$=1, 2, 3, 5, 7, 8 
and 10 can be represented as ($\mu_x-1$) $\mu_B$, whereas for $x$=4 and 9, it can be expressed as 
($\mu_x-3$) $\mu_B$, where $\mu_x$ is the total magnetic moment of the Mn$_x$ cluster corresponding to 
the ground state or the first isomer \cite{kabir}.   
Local magnetic moment, $M$, at each site can be calculated by 
$M = \int_0^R[\rho_{\uparrow}(\mathbf r) - \rho_{\downarrow}(\mathbf r)] d\mathbf r$, where 
$\rho_{\uparrow}(\mathbf r)$ and $\rho_{\downarrow}(\mathbf r)$ are up-spin and down-spin charge densities,
respectively, and $R$ \cite{localmag} is the radius of a sphere centring the atom. For MnAs dimer, 
the magnetic moment at Mn site, $M_{Mn}$, and at As site, $M_{As}$, are 3.72 $\mu_B$ and  $-$0.26 $\mu_B$,  
respectively.  
This large negative polarization of the anion, As, is due to the strong $p-d$ interaction. In the Mn$_2$As
cluster, Mn atoms are ferromagnetically coupled with $M_{Mn}$=3.79 $\mu_B$ each, whereas $M_{As}$ is
 $-$0.14 $\mu_B$. Mn atoms in the Mn$_3$As take frustrated antiferromagnetic structure with 
$M_{Mn} = $ 3.1, 3.1 and $-$3.9 $\mu_B$ and $M_{As}$ has a value $-$0.21 $\mu_B$.
For Mn$_4$As, Mn atoms are ferromagnetically arranged (Fig.\ref{fig:densities})
with an average $M_{Mn}$=3.66 $\mu_B$ and are coupled antiferromagnetically with the As atom, $M_{As}$=$-$0.22 
$\mu_B$. 
For Mn$_5$As, $M_{Mn}$ varies between 3.04-3.72 $\mu_B$ with a local magnetic moment $-$0.23 
$\mu_B$ at As site. 
Polarized neutron diffraction study found a local magnetic moment of $-$0.23 $\pm$ 0.05 $\mu_B$ at the 
As sites for NiAs-type MnAs \cite{yam}, which is very close to the present values for MnAs - Mn$_5$As clusters.
These Mn-Mn magnetic behavior are unlike the previous study of nitrogen-doped Mn$_x$ clusters,
$x$=1-5, by Rao and Jena \cite{rao}, where Mn atoms were found to be coupled ferromagnetically for all sizes. 
We observe, the negative polarization of As, $M_{As}$, decreases sharply to 
$-0.08$ $\mu_B$ for Mn$_6$As and further decreases monotonically to $-$ 0.02 $\mu_B$ for Mn$_{8}$As, 
however it becomes positive, 0.04 and 0.02, for Mn$_9$As and Mn$_{10}$As, respectively. 
For all Mn$_x$As clusters, $x=$6-10, Mn atoms are coupled ferrimagnetically, and $M_{Mn}$ of highly 
coordinated atoms are very small than those of the surface atoms and it varies between 0.8 - 3.7 $\mu_B$.  

In conclusion, though the As atom induces clustering of Mn, the individual magnetic moments of Mn 
couple ferromagnetically only for Mn$_2$As and Mn$_4$As clusters and, however, ferrimagnetically aligned
for all other sizes. Not only that, the ferromagnetic Mn$_2$As and Mn$_4$As are even more stable than
other sized ferrimagnetic clusters, as they have largest spin gaps.
We believe that Mn clustering, during the low temperature MBE growth, could 
be responsible for the ferromagnetism and reported high $T_c$ in (GaMn)As and should be taken 
into account to formulate an adequate theory of ferromagnetism in III-V semiconductors \cite{schil}. 
If Mn-doped GaAs, under carefully controlled growth and annealing condition, contains Mn$_2$As and/or 
Mn$_4$As clusters, which have large cluster magnetic moments, would enhance the local magnetic moment and
consequently could enhance $T_c$. On the other hand, large clustering would yield low $T_c$ due
to their small local magnetic moment. This study also suggest that the similar mechanism could enhance 
the $T_c$ of (InMn)As and further investigation of Mn-clustering around oxygen could explain the 
ferromagnetism and observed high $T_c$ of Mn-doped ZnO in the {\it metastable} phase \cite{nature}.
EXAFS studies would be very useful to see whether these clusters of Mn around As are, indeed, present
in these samples and the gas phase experiments involving Mn clustering in a As-seeded chamber 
can yield direct information on the magnetic behavior of Mn$_x$As clusters. We hope, our study 
will encourage such experiments. 

M. K. thankfully acknowledges the congenial hospitality at the Centre for Modelling and Simulation of Pune
University. This work has been done under the DST contract SR/S2/CMP-25/2003.

\vskip 4cm 
{\centerline {Figure Captions}}
1. Ground state geometries and the corresponding
spin ordering of Mn$_x$As
clusters, $x$=1-10. Blue and red balls represent Mn$_{\uparrow}$ and Mn$_{\downarrow}$ atoms,
respectively. Green ball represents the As atom. The bond lengths are given in \r{A}. Magnetic
polarization of As, is negative for MnAs - Mn$_8$As and, whereas, positive for Mn$_9$As and Mn$_{10}$As.
Note, Mn-Mn interaction is ferromagnetic only for Mn$_2$As and Mn$_{4}$As clusters and ferrimagnetic
for all other sizes. \\

2. (a) Plot of binding energy/atom with $x$, for Mn$_x$ and Mn$_x$As.
(b) Two energy gains, $\Delta^1$ and $\Delta^2$, are plotted with $x$. These energy gains are defined as,
$\Delta^1$ = $-$ [$E$(Mn$_x$As) $-$ $E$(Mn$_x$) $-$ $E$(As)] and $\Delta^2$ = $-$ [$E$(Mn$_x$As) $-$
$E$(Mn$_{1-x}$As) $-$ $E$(Mn)]. \\

3. Constant spin density surfaces  for Mn$_2$As, Mn$_3$As,
 Mn$_4$As and Mn$_8$As
corresponding to 0.04, 0.04, 0.04 and 0.02 $e$/\r{A}$^3$, respectively. Red
and blue surfaces represent positive and negative spin densities, respectively. Green ball is
the As atom, which has negative polarization in all these structures. Note ferromagnetic
(Mn$_2$As and Mn$_4$As (left panel)) and ferrimagnetic (Mn$_3$As and Mn$_8$As (right panel)) coupling
between Mn atoms.


\begin{thebibliography}{99}
\bibitem{localmag} This procedure captures at least 85\% of the total moment for the clusters. $R_{Mn}$ is taken 
1.10 \r{A} for all $x$, whereas, $R_{As}$ are 1.11, 1.22 and 1.30 \r{A} for $x$=1,2 and 3-10, respectively.
\end{thebibliography}
\end{document}